\documentclass{ws-procs9x6-cpt19}
\usepackage[english]{babel}
\usepackage[utf8]{inputenc}
\usepackage{lineno}
\usepackage{arydshln}
\usepackage{xspace}
\def\LV{\ifmmode {\mathrm{LV}}\else{\scshape LV}\fi\xspace}
\def\LIV{\ifmmode {\mathrm{LV}}\else{\scshape LV}\fi\xspace}
\def\LI{\ifmmode {\mathrm{LI}}\else{\scshape LI}\fi\xspace}

\newcommand{\refeq}[1]{(\ref{#1})}

\begin{document}


\title{Lorentz violation constraints with astroparticle physics
}

\author{H.\ Mart\'inez-Huerta\footnote{The author acknowledges FAPESP support No. 2015/15897-1 and 2017/03680-3 and the National Laboratory for Scientific Computing (LNCC/MCTI, Brazil) for providing HPC resources of the SDumont supercomputer, which have contributed to the research results reported within this paper (http://sdumont.lncc.br).}
}

\address{
Instituto de F\'isica de S\~ao Carlos, Universidade de S\~ao Paulo,\\
Av. Trabalhador S\~ao-carlense 400, CEP 13566-590 S\~ao Carlos, SP, Brasil
}

\begin{abstract}
Astroparticle physics has recently reached a new status of precision due to the construction of new observatories, operating innovative technologies and the detection of large numbers of events and sources. The precise measurements of cosmic and gamma rays can be used as a test for fundamental physics, such as the Lorentz invariance violation (\LIV). Although \LIV signatures are expected to be small, the very high energies and long distances that astrophysical sources involve, lead to an unprecedented opportunity for this task. In this summary, exclusion limits results are presented from different types of astrophysical \LV tests through the generic modification of the particle dispersion relation in the photon sector through the pair production threshold shifts and photon decay. Some perspectives for the next generations of gamma-ray telescopes are also addressed.
\end{abstract}

\bodymatter

\section{MDR for astroparticle tests}

The introduction of a Lorentz violating term in the Lagrangian \cite{Coleman} or spontaneous Lorentz symmetry breaking \cite{Colladay:1998fq} can induce modifications to the particle dispersion relation (MDR). A phenomenological generalization of the \LV effects converges on the introduction of a general function of the energy and the momentum. Although, there are several forms of MDR for different particles and underlying \LV -theories, some of them may lead to similar phenomenology, which can be useful for \LV tests in extreme environments such as astroparticle scenarios. In this line of thought, a family of MDRs can be addressed by the following expression in natural units,
\begin{equation}\label{eq:GDR}
    E_{a}^{2} -  p_{a}^{2} = m_a^2 \pm |\delta_{a,n}|A_a^{n+2},
\end{equation}
where $a$ stands for the particle type. $A$ can take the form of E or p. $\delta_{a,n}$ is the \LV parameter and $n$, is the leading order of the correction from the underlying theory. In some effective field theories, $\delta_{a,n}=\epsilon^{(n)}/M$, where $M$ is the energy scale of the new physics, such as, the Plank energy scale, $\rm E_{Pl}$, or some Quantum Gravity energy scale, $\rm E_{QG}$, and $\epsilon^{(n)}$ are \LV coefficients.

\section{Gamma-ray attenuation including \LV}

Very-high gamma rays that propagate from distant sources suffer significant attenuation due to the interaction with the background light (BL) through the pair production process
$\gamma~\gamma_{b} \longrightarrow  e^+ e^-,$
which constrains how far in the universe we can expect photons without being absorbed\cite{DeAngelis:2013jna}. 
Indices $b$ and $\pm$ denote BL and $e^\pm$.
The derived physics from Eq.~\refeq{eq:GDR} can leads to shifts at the minimum BL energy that allows the pair production process~\cite{Martinez-Huerta:2016azo}, given by 
\begin{equation}\label{eq:TH1}
E_{\gamma_b}^{th} = \frac{m_e^2}{4E_{\gamma} K(1-K)} - \frac{1}{4} \delta^{tot}_n \ E_{\gamma}^{n+1},
\end{equation}
where  $\delta^{\rm tot}_n =  \delta_{\gamma,n} - \delta_{+,n}K^{n+1} - \delta_{-,n}(1-K)^{n+1}$, is a linear combination of the \LIV coefficients from the different particle species  and  $K$ stands for the inelasticity\cite{Martinez-Huerta:2016azo,BHCB}, $E_{+}= K (E_\gamma + E_{\gamma_b})$. 
The cumulative effect of this phenomenon results in measurable changes in the expected attenuation of the gamma-rays flux due to the BL~\cite{TeV-LV,UHECR-LV}. 
Due to its nature in the universe, dominant BL in different gamma-ray energy windows should be considered, for instance, for $E_\gamma < 10^{14.5}$ eV, the extragalactic background light (EBL) is the dominant BL, and for $10^{14.5} \mathrm{eV} < E_\gamma < 10^{19}$eV, is the cosmic background microwave radiation (CMB).
In the EBL energy region, the subluminal ($\delta_{n}^{tot}<0$) \LV effect forecast a recovery in the spectrum of TeV-sources that can be measured by the current gamma-ray telescopes~\cite{TeV-limits}. 
Ref.~\refcite{TeV-LV} reported a new analysis procedure to search this LV-signature with the most updated TeV gamma-ray data set, by looking at 111 measured energy spectra from 38 sources. It was concluded that the data set is best described by the LI model and the stringent limits at 2 sig. CL in the Tab.~\ref{tab:1} (Ref. \refcite{TeV-LV}), 
were established, which, by first time in the literature, are robust under several tested systematic uncertainties.

The Cherenkov Telescope Array (CTA) is the next generation ground-based IACT observatory for gamma-ray astronomy at very-high energies. It will be capable of detecting gamma rays in the energy range from 20 GeV to more than 300 TeV with unprecedented precision in energy and directional reconstruction, which will generate an unparalleled opportunity for LV-tests\cite{CTA-KS}. Preliminary results from the simulations of CTA observations of the nearby blazar Mrk 501 spectrum have shown that CTA will be sensitive to this type of signatures with at least the \LV values in Tab.\ref{tab:1} (Ref. \refcite{CTA-ICRC}).

Ref.~\refcite{UHECR-LV} performed a search for  the \LV signatures in the CMB dominant background region, by computing the so-called GZK-photon flux on Earth and considering for the first time in the literature, several ultra-high-energy cosmic rays (UHECR) injection and source distribution models, including the model combination that was shown to best describe the energy spectrum, composition, and arrival direction of UHECR data, which corresponds to a source distribution model that follows a GRB rate evolution, with a power-law energy spectrum injection model at the source with a rigidity cutoff, and including five different species of primary cosmic ray nuclei\cite{UHECR}. 
The resulting \LV effect was the increase in the predicted GZK-photon flux\cite{BHCB,UHECR-LV}. Then, by comparing these results with the most updated upper limits to the integrated photon flux obtained by the Pierre Auger Observatory\cite{AugerLimits}, the \LIV-scenarios in Tab.\ref{tab:1} (Ref. \refcite{UHECR-LV}), were excluded, since they predict GZK-photons above the current upper photon limits, in the astrophysical scenario which best describes UHECR data.

\section{Photon decay}

The phenomenology derived from Eq.~\refeq{eq:GDR}, can also study superluminal phenomena predicted in some \LIV scenarios such as the photon decay, photons splitting and  vacuum Cherenkov radiation\cite{Martinez-Huerta:2016azo,PhS}. 
The resulting decay rates for the photon decay into electron-positron are very fast and effective once the process is allowed, which suggests an abrupt cutoff in the gamma-ray spectrum with no high-energy photons reaching the Earth from cos\-mo\-lo\-gi\-cal distances above a given threshold. The threshold for any order~$n$ from Eq.~\refeq{eq:GDR}, is given by~\cite{Martinez-Huerta:2016azo} $\delta_{\gamma,n} \ge 4 m^2 /(E_{\gamma}^{n}(E_{\gamma}^2-4 m^2))$.
Using the reported $E_{\gamma}=56$ TeV observations at 5 sig. of CL from the Crab Nebula, by the HEGRA telescope, the exclusion limits in Tab.\ref{tab:1}~(Ref.~\refcite{Martinez-Huerta:2016azo}) were reported.

The very recent results of gamma-ray energies above 100 TeV by the High Altitude Water Cherenkov (HAWC) observatory~\cite{HAWC}, together with recent development of an energy-reconstruction algorithm, new and stringent constraints are being established on the order of those in Tab.\ref{tab:1}~(Ref.~\refcite{HAWC-LV}), see J.T. Linnemann for the HAWC Collaboration, in this Proceedings. 

In addition, prospects to test photon decay through the observation of the SNR RX J1713.7-394 were reported in the science motivation paper for the Southern Gamma-Ray Survey Observatory (SGSO), which will be a next-generation wide field-of-view gamma-ray survey instrument, sensitive to gamma-rays in the energy range from 100 GeV to hundreds of TeV, which can lead to exclusion limits on the order of those in Tab.\ref{tab:1}~(Ref.~\refcite{SGSO}).

\begin{table}
\tbl{\LV limits. PP: pair production. PD: photon decay.\label{tab:1}}
{
\begin{tabular}{@{}lcccccll@{}} \toprule
\multicolumn{1}{c}{Type} & \begin{tabular}[c]{@{}c@{}}$|\delta_0|$\\  $10^{-20}$\end{tabular} & \begin{tabular}[c]{@{}c@{}}$|\delta_1|$ \\ $10^{-30}$eV$^{-1}$\end{tabular} & \begin{tabular}[c]{@{}c@{}}$|\delta_2|$\\ $10^{-43}$eV$^{-2}$\end{tabular} & \begin{tabular}[c]{@{}c@{}}$\rm E_{\LV}^{(1)}$\\  $10^{28}$eV\end{tabular} & \begin{tabular}[c]{@{}c@{}}$\rm E_{\LV}^{(2)}$\\ $10^{21}$eV\end{tabular} & \multicolumn{1}{c}{\begin{tabular}[c]{@{}c@{}}Threshold\\ bound\end{tabular}} & Ref. \\ \toprule
Limit & - & 8.3 & 1.8 & 12.08 & 2.38 & \begin{tabular}[c]{@{}l@{}}PP ($\delta^{tot}_{n}<0$)\end{tabular} & [\refcite{TeV-LV}] \\
Limit$^\dagger$ & $\sim$ 1 & $\sim 10^{-8}$ & $\sim 10^{-13}$ & $\sim 10^{10}$ & $\sim 10^{7}$ & \begin{tabular}[c]{@{}l@{}}PP ($\delta^{tot}_{n}<0$)\end{tabular} & [\refcite{UHECR-LV}] \\ 
\begin{tabular}[c]{@{}l@{}}
CTA  Sens.lim.2017\end{tabular} & - & $\sim$82 & $\sim$11 & $\sim$1.22 & $\sim$0.97 & \begin{tabular}[c]{@{}l@{}}PP ($\delta^{tot}_{n}<0$)\end{tabular} & [\refcite{CTA-ICRC}] \\ 
Limit & - & 6.7 & 1.3 & 15.0 & 2.8 & \begin{tabular}[c]{@{}l@{}}PD ($\delta_{\gamma, n}>0$)\end{tabular} & [\refcite{Martinez-Huerta:2016azo}] \\
\begin{tabular}[c]{@{}l@{}}
HAWC limit (Prel.)\end{tabular} 
& - & $\lesssim 0.1$ & $\lesssim 0.01 $& $\gtrsim 10^{3}$ & $\gtrsim 10$ &
\begin{tabular}[c]{@{}l@{}}PD ($\delta_{\gamma, n}>0$)\end{tabular} & [\refcite{HAWC-LV}] \\
\begin{tabular}[c]{@{}l@{}}
SGSO Sens.lim.2019\end{tabular} & - & $\sim 1$ & $\sim 0.1$ & $\sim 10^{2}$ & $\sim 10$ & \begin{tabular}[c]{@{}l@{}}PD ($\delta_{\gamma, n}>0$)\end{tabular} & [\refcite{SGSO}]
\\ \botrule
\end{tabular}
}
\begin{tabnote}
$\delta_{n}$ $=(\rm E_{\rm LV}^{(n)})^{-n} \approx \rm (\rm E_{\rm QG}^{(n)})^{-n}$. $\rm E_{\rm Pl} \approx 1.22 \times 10^{28}$eV. \\ $\dagger$ Limits in the astrophysical scenario which best describes UHECR data.
\end{tabnote}
\end{table}

\section{Conclusions}

The precise measurements of cosmic and gamma rays can be used to test for fundamental physics, such as the Lorentz violation. New exclusion limits derived from the effect of shifting the energy threshold of pair production and the instability of photon, compatible with updated data of TeV sources, UHECR, and gamma-rays, were presented. Updates and new studies can be expected with the advent of new and better data from the cosmic messengers provided by the current experiments as HAWC and the next-generation such as the Cherenkov Telescope Array and SGSO. 


\end{document}